# Coherent Perfect Tunneling at Exceptional Points via Directional Degeneracy

Huayang Cai[a,b,c] and Bishuang Chen[d, *]

Author affiliations: [a]Institute of Estuarine and Coastal Research, School of Ocean Engineering and Technology, Sun Yat-Sen University / Southern Marine Science and Engineering Guangdong Laboratory (Zhuhai), Zhuhai, Guangdong 519082, China; [b]State and Local Joint Engineering Laboratory of Estuarine Hydraulic Technology / Guangdong Provincial Engineering Research Center of Coasts, Islands and Reefs / Guangdong Provincial Key Laboratory of Marine Resources and Coastal Engineering / Guangdong Provincial Key Laboratory of Information Technology for Deep Water Acoustics / Key Laboratory of Comprehensive Observation of Polar Environment (Sun Yat-Sen University), Ministry of Education, Zhuhai, Guangdong 519082, China; [c]Zhuhai Research Center, Hanjiang National Laboratory, Zhuhai, Guangdong 519082, China; [d]School of Marine Sciences, Sun Yat-Sen University, Zhuhai 519080, China.

[*]Corresponding author: Bishuang Chen.

**Email:** chenbsh23@mail.sysu.edu.cn

## Abstract

Coherent perfect tunneling in the presence of loss and asymmetry remains a fundamental challenge in wave transport, a universal problem across optics, acoustics, and quantum mechanics. Here we demonstrate coherent perfect tunneling at an exceptional point in a passive one-dimensional waveguide cascade with three coupled interfaces. Using a waveguide-invariant scattering framework, we show that the suppression of a selected output channel originates from a directional scattering degeneracy rather than from resonance or absorption collapse. This exceptional-point condition emerges when interference between boundary-induced feedback loops promotes a simple zero of the scattering response to a second-order degeneracy. As a direct consequence, fixed coherent excitation produces a robust quartic leakage law within a transparency-dominated tunneling window. These results establish directional degeneracy as a general mechanism for loss-tolerant tunneling enabled by exceptional points across a broad class of wave systems.

## 1. Introduction

Wave tunneling through finite, lossy structures underpins energy and information transport across optics, acoustics, quantum systems, and electrochemical interfaces. In conventional descriptions, tunneling and transmission are governed by resonance conditions, impedance matching, or modal hybridization, concepts that trace back to classical wave and mode theories (1). While such approaches successfully explain resonant tunneling and critical coupling in idealized settings, they remain intrinsically sensitive to loss, asymmetry, and boundary detuning, making perfect transmission fragile in realistic non-Hermitian systems (2, 3). Beyond photonics, the same boundary-controlled transport challenge recurs in acoustic cloaking and metasonics, microwave networks, and quantum-dot or waveguide-quantum electrodynamics junctions, where loss and fabrication asymmetry are unavoidable. Existing exceptional-point concepts have largely been deployed for absorption control and sensing; here we show how an exceptional point can instead protect useful power delivery, enabling near-unitary tunneling while suppressing a selected outgoing channel.

Recent advances in exceptional-point (EP) physics have demonstrated that non-Hermitian degeneracies can qualitatively reshape wave transport and interference. In optics, coherent perfect absorption (CPA) at EPs was shown to exhibit anomalous quartic spectral flattening and enhanced robustness (4), and was subsequently extended to massively degenerate and wavefront-agnostic configurations (5, 6), time-varying media with simultaneous absorption and amplification (7), and even quantum states of light (8). These developments establish EPs as a powerful mechanism for controlling dissipation and transport. However, existing EP-based formulations of tunneling and absorption are predominantly expressed in terms of scattering-matrix eigenmodes or coupled-mode models (9), which obscure the energetic distinction between total dissipation and useful transmitted power, and offer limited insight into how directionality and asymmetry fundamentally constrain perfect tunneling.

Here we show that coherent perfect tunneling (CPT) at exceptional points emerges naturally as a directional degeneracy within a waveguide-invariant scattering framework. Building on a recently established universal mass–energy relation for lossy one-dimensional waveguides (10), and its kinetic extension to electrochemical polarization (11), we demonstrate that CPT–EP conditions correspond to a geometric collapse of the effective standing-wave magnitude along a specific transport direction, rather than to conventional resonance. This invariant description reveals CPT as a boundary-controlled, direction-selective degeneracy in energy–power space, unifying tunneling, exceptional points, and useful power delivery under a common framework that remains valid in the presence of intrinsic asymmetry and distributed loss.

## 2. Results

### 2.1. Waveguide-invariant scattering with three coupled interfaces

We consider a power-normalized two-port scattering system $\mathbf{b}(\Omega) = \mathbf{S}(\Omega)\mathbf{a}$, where $\mathbf{a} = (a_1, a_2)^T$ and $\mathbf{b} = (b_1, b_2)^T$ denote the complex input and output wave amplitudes, respectively, and $\Omega$ is the dimensionless angular frequency. The structure consists of a passive one-dimensional waveguide cascade formed by two identical lossy propagation segments separated by three lossless interfaces characterized by reflection amplitudes $\Gamma_S$, $\Gamma_I$, and $\Gamma_L$, corresponding to the source, internal, and load interfaces. Distributed attenuation and phase accumulation are encoded in the complex propagation constant $K(\Omega) = \sqrt{(i\Omega + \delta_R)(i\Omega + \delta_G)}$, where $\delta_R$ and $\delta_G$ are non-negative loss parameters associated with the two propagation directions. The resulting round-trip feedback factor is $z(\Omega) = \exp[-2K(\Omega)]$. In this formulation, all frequency dependence enters exclusively through $z(\Omega)$, while the interface reflections control the coupling between multiple scattering paths. This separation renders the description waveguide-invariant: geometric length, loss, and dispersion are absorbed into a single complex scalar $z$, whereas boundary-induced interference is governed solely by the interface coefficients.

### 2.2. Analytic numerator–denominator structure and CPT–EP condition

For a fixed excitation direction, the complex amplitude of the suppressed output channel can be expressed in the universal rational form

$$b_1(\Omega) = \frac{N[z(\Omega)]}{D[z(\Omega)]}, \tag{1}$$

where the denominator $D(z)$ originates from the total ABCD transfer matrix of the cascade and remains finite in the tunneling-dominated regime. The numerator explicitly reads

$$N(z) = (A + B - C - D)a_1 + 2(AD - BC)a_2, \tag{2}$$

with $A, B, C, D$ denoting the ABCD elements of the full three-interface cascade, each being analytic functions of $z$ determined by the mutual coupling of $\Gamma_S$, $\Gamma_I$, and $\Gamma_L$ (see *SI Appendix* for the full ABCD derivation). With only two interfaces, the numerator contains at most a single controllable interference loop and generically supports only a simple zero, yielding linear suppression in complex amplitude. The internal interface introduces an additional independent feedback path, which is the minimal degree of freedom needed to promote a simple zero to a double zero under passive constraints, thereby enabling CPT–EP. In the fixed coherent operation protocol we hold the incident state $\mathbf{a} = \mathbf{a}_{\text{fix}}$ constant while sweeping $\Omega$. Consequently $N$ becomes an analytic scalar function of $z(\Omega)$ alone, and the CPT–EP condition is imposed as

$$N\left[z(\Omega_0); \mathbf{a}_{\text{fix}}\right] = 0, \left.\frac{dN}{dz}\right|_{z(\Omega_0), \mathbf{a}_{\text{fix}}} = 0, D\left[z(\Omega_0)\right] \neq 0. \tag{3}$$

By contrast, in the per-frequency probe protocol one may choose $\mathbf{a} = \mathbf{a}_{\text{pf}}(\Omega)$ such that $b_1(\Omega) = 0$ identically, which does not require the derivative constraint in Eq. (3). This double-root condition defines a directional scattering degeneracy: a selected outgoing channel is suppressed without inducing resonance or absorption collapse. Analytically, the exceptional-point closure of the feedback factor is given by

$$z_{\text{EP}} = -\sqrt{\Gamma_S/\Gamma_L}, \tag{4}$$

which fixes $\Omega_0$ through $z(\Omega_0) = z_{\text{EP}}$. The closure (4) has a transparent physical meaning: it enforces a geometric-mean balance between the source and load boundary feedback loops, while the minus sign fixes the relative phase required for destructive interference in the selected output channel under the chosen phase gauge. The internal interface must simultaneously satisfy the critical coupling condition

$$|\Gamma_I| = \frac{2\sqrt{|\Gamma_S\Gamma_L|}}{1+|\Gamma_S\Gamma_L|}, \tag{5}$$

balancing the competing feedback loops generated by the source and load interfaces. Without the internal interface, the system supports at most a simple zero in $N(z)$, leading only to linear suppression of the complex amplitude; the third interface provides the additional interference degree of freedom required to promote this simple zero to a second-order degeneracy under passive conditions.

### 2.3. Near-unitary tunneling and quartic leakage at CPT–EP

**Figure 1** visualizes the physical consequences of the CPT–EP condition. In **Fig. 1A**, both the reflection coefficient $|S_{11}|$ and the fixed-input throughput $\|\mathbf{Sa}_{\text{fixed}}\|$ remain close to unity around $\Omega_0$, while deviations from unitarity, quantified by $1 - \sigma_{\min}$ and $1 - \sigma_{\max}$, stay below $10^{-4}$, confirming that CPT–EP operates in a transparency-dominated tunneling window. **Figure 1B** shows the coherent input ratio $a_2/a_1$ required to suppress the output channel: under per-frequency probing it varies rapidly with $\Omega$, whereas under fixed excitation it remains constant, isolating intrinsic scattering sensitivity. **Figure 1C** presents the power balance, demonstrating that the suppressed-channel power $P_1$ vanishes at $\Omega_0$ while the transmitted power $P_2$ remains near unity, clearly distinguishing CPT–EP from coherent perfect absorption. Finally, **Fig. 1D** reveals the hallmark CPT–EP scaling: for fixed coherent excitation, the suppressed-channel leakage obeys the quartic law $P_{1,\text{fixed}} \propto |\Omega - \Omega_0|^4$, whereas per-frequency probing trivially enforces $P_1 = 0$. This quartic suppression is the direct experimental signature of the second-order zero of the waveguide-invariant numerator $N(z)$ and confirms the exceptional-point nature of coherent perfect tunneling. Together, these four panels establish that CPT–EP is neither a resonant nor an absorptive phenomenon, but a new class of transport degeneracy: a direction-selective exceptional-point collapse that suppresses one output channel while preserving near-unitary throughput.

It is worth noting that the absolute phases of $\Gamma_S$, $\Gamma_I$, $\Gamma_L$ (denoted by $\phi_S$, $\phi_I$, and $\phi_L$, respectively) are not all physically independent: a global phase rotation of the field amplitudes leaves measurable powers invariant. We therefore fix a convenient gauge by choosing $\phi_S = 0$, $\phi_I = 0$, $\phi_L = \pi$, so that $\Gamma_S$, $\Gamma_I$ are taken real and positive while $\Gamma_L$ is real and negative. This choice simplifies the closed-form expressions (e.g., $z_{\text{EP}}$) without loss of generality; other phase conventions lead to equivalent conditions obtained by a corresponding rephasing of **a**.

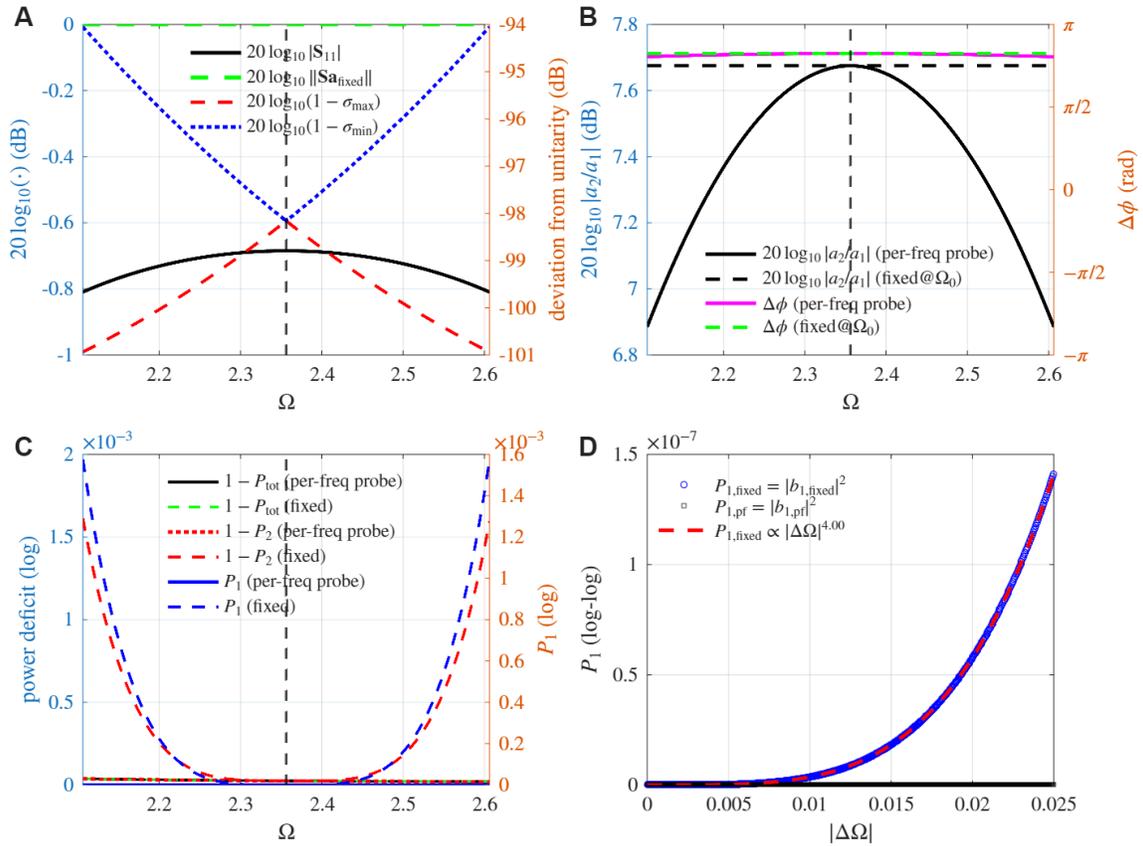

**Fig. 1. Coherent perfect tunneling at an exceptional point (CPT–EP) in a passive ABCD waveguide cascade: per-frequency probe versus fixed coherent operation.**

**A,** Near-unitary tunneling window around $\Omega_0$ (vertical dashed line): $20\log_{10}|S_{11}|$ and the fixed-input throughput $20\log_{10}\|\mathbf{Sa}_{\text{fixed}}\|$ (left axis), together with deviations from unitarity $20\log_{10}(1-\sigma_{\max})$ and $20\log_{10}(1-\sigma_{\min})$ (right axis), where $\sigma_{\max,\min}$ are singular values of $\mathbf{S}(\Omega)$. **B,** Required coherent input ratio and phase, shown as $20\log_{10}|a_2/a_1|$ and $\Delta\phi = \arg(a_2/a_1)$ for per-frequency probing (solid) and the fixed operating state (dashed). **C,** Power-balance diagnostics in log scale: the total power deficit $1-P_{\text{tot}}$ and the transmitted-channel deficit $1-P_2$ (left axis), together with the suppressed-channel power $P_1 = |b_1|^2$ (right axis), for both protocols; CPT–EP manifests without absorption collapse, maintaining $P_2 \approx 1$ while suppressing $P_1$ near $\Omega_0$. **D,** Quartic leakage law under fixed coherent operation: on the detuning grid $\Delta\Omega = \Omega - \Omega_0 > 0$, the fixed-input leakage follows $P_{1,\text{fixed}} \propto |\Delta\Omega|^4$, whereas the per-frequency probe remains at numerical floor because $b_1$ is nulled by construction.

## 3. Discussion and Conclusion

This work establishes CPT–EP as a distinct non-Hermitian transport regime that is fundamentally different from both resonant tunneling and coherent perfect absorption. In CPT–EP, the suppression of a selected output channel arises from a directional scattering degeneracy rather than from resonance-induced field enhancement or absorption collapse, enabling near-unitary transmission to be maintained. A key insight is the essential role of the internal interface. Systems with only source and load boundaries support at most a simple zero of the scattering response, leading to linear suppression of the complex amplitude. Introducing a third interface provides an additional interference pathway that promotes this simple zero to a second-order degeneracy under passive conditions, thereby stabilizing the EP against loss and asymmetry.

The observable hallmark of this degeneracy is the quartic leakage law under fixed coherent excitation. Unlike per-frequency optimized control, where suppression is enforced trivially by adaptive inputs, the quartic scaling reflects an intrinsic property of the scattering structure and offers a robust experimental signature of CPT–EP.

More broadly, the waveguide-invariant formulation adopted here shows that exceptional-point-enabled tunneling can be understood in terms of boundary-induced interference acting on a single feedback variable, independent of microscopic geometry. This perspective identifies directional degeneracy as a general and loss-tolerant mechanism for controlling wave transport, with potential implications for photonic, acoustic, and quantum systems where dissipation and boundary asymmetry are unavoidable.

**Materials and Methods**

We model CPT–EPs using a waveguide-invariant scattering framework. The system consists of a passive one-dimensional waveguide cascade with three coupled interfaces, where feedback loops and interface reflection coefficients are key. The numerator and denominator of the output scattering response are derived analytically, and the CPT–EP condition is enforced by solving a set of closure equations and critical coupling conditions. The methods also include fixed coherent excitation and per-frequency probing protocols to explore tunneling behavior across different regimes, with power leakage quantified by a quartic law near the EP. Detailed methods can be found in the *SI Appendix*.

**Data, Materials, and Software Availability.** The MATLAB scripts used for reproducing all figures are openly released at https://github.com/Huayangcai/Coherent-Perfect-Tunneling-at-Exceptional-Points.

## Acknowledgments

This research was supported by the Guangdong Basic and Applied Basic Research Foundation (Grant No. 2023B1515040028), and the National Natural Science Foundation of China (Grant No. 52279080, 42376097).

**Supporting Information for**

**Coherent Perfect Tunneling at Exceptional Points via Directional Degeneracy**

Huayang Cai[a,b,c] and Bishuang Chen[d, *]

*Corresponding author: Bishuang Chen.

**Email:** chenbsh23@mail.sysu.edu.cn

**Supporting Information Text**

**Extended Methods**

**S1. Waveguide cascade model, interface coefficients, and $\mathbf{S}(\Omega)$ construction**

We consider the normalized two-port scattering system $\mathbf{b}(\Omega) = \mathbf{S}(\Omega)\mathbf{a}(\Omega)$, with $\mathbf{a} = (a_1, a_2)^T$, $\mathbf{b} = (b_1, b_2)^T$, and output powers $P_k = |b_k|^2$ (matched reference basis, equal port impedances). The $2 \times 2$ scattering matrix is given by

$$\mathbf{S}(\Omega) = \begin{pmatrix} \mathbf{S}_{11}(\Omega) & \mathbf{S}_{12}(\Omega) \\ \mathbf{S}_{21}(\Omega) & \mathbf{S}_{22}(\Omega) \end{pmatrix}. \tag{S1}$$

In a system with only two interfaces (source and load), the frequency dependence of any fixed-output channel is governed by a single feedback loop, and the corresponding channel numerator generally possesses only a simple zero; under passive distributed loss, the best achievable suppression near a tuned frequency is linear in the complex amplitude. Introducing an internal interface creates a second, controllable interference loop, which promotes a simple zero to a double zero in the channel numerator while maintaining nearly unitary global scattering. This additional degree of freedom is the minimal mechanism required to realize CPT–EP: a direction-selective, EP-protected quartic leakage law within a transparency-dominated window.

Each propagation section is modeled by a symmetric ABCD matrix

$$\mathbf{T_p}(\Omega) = \begin{pmatrix} \cosh K(\Omega) & \sinh K(\Omega) \\ \sinh K(\Omega) & \cosh K(\Omega) \end{pmatrix}, \quad (S2)$$

where the complex propagation constant is

$$K(\Omega) = \sqrt{(i\Omega + \delta_R)(i\Omega + \delta_G)}, \delta_R, \delta_G \geq 0, \quad (S3)$$

and the square-root branch is chosen so that $\text{Re}[K(\Omega)] \geq 0$ for all sampled $\Omega$. We compress distributed propagation and loss into the single scalar (i.e., one-way feedback factor)

$$z(\Omega) = e^{-2K(\Omega)}. \quad (S4)$$

If the physical device contains two identical segments, the corresponding transfer matrix naturally generates both $z(\Omega)$ and $z^2(\Omega)$ terms; equivalently, one may absorb the two-segment length into $K(\Omega)$ so that $z(\Omega)$ represents the full interfacial separation. Throughout this work we adopt the former (segment-wise) convention unless stated otherwise. The structure contains three interfaces with reflection coefficients $\Gamma_S, \Gamma_I, \Gamma_L$, where $0 \leq |\Gamma_I| < 1$. Each is parametrized as $\Gamma_j = |\Gamma_j|e^{i\phi_j}$. For the implementation used in Fig. 1, the phases are fixed as

$$\phi_S = 0, \phi_I = 0, \phi_L = \pi, \quad (S5)$$

so that $\Gamma_S = |\Gamma_S|$, $\Gamma_I = |\Gamma_I|$, and $\Gamma_L = -|\Gamma_L|$. Each interface is modeled as a symmetric unitary two-port with transmission amplitude

$$\tau_j = \sqrt{1 - |\Gamma_j|^2}, \quad (S6)$$

and scattering matrix

$$\mathbf{S}_{\Gamma_j} = \begin{pmatrix} \Gamma_j & i\tau_j \\ i\tau_j & \Gamma_j \end{pmatrix}. \quad (S7)$$

The imaginary unit in the off-diagonal terms is a phase-gauge choice enforced by unitarity.

For a lossless symmetric two-port $\mathbf{S} = \begin{pmatrix} r & t \\ t & r \end{pmatrix}$, the condition $\mathbf{S}^\dagger \mathbf{S} = \mathbf{I}$ implies $|r|^2 + |t|^2 = 1$ and $r^*t + t^*r = 0$, i.e., $\text{Re}(r^*t) = 0$. Hence $t$ must be $\pm\pi/2$ out of phase with $r$. Writing $r = \Gamma_j$ and $t = i\tau_j$ with $\tau_j = \sqrt{1 - |\Gamma_j|^2}$ satisfies these constraints, and $\pm i\tau_j$ are equivalent up to a trivial port-phase redefinition.

Following the waveguide-invariant formulation of ref. 10, the distributed dynamics enter only through the scalar $z(\Omega) = e^{-2K(\Omega)}$, while boundary control is encoded by a Möbius transformation that composes the source and load reflections into an effective generalized reflection,

$$\bar{\Gamma}_g(\Omega) = \frac{\Gamma_L\, z(\Omega) + \Gamma_S}{1 + \Gamma_S \Gamma_L\, z(\Omega)}. \quad (S8)$$

This mapping provides a low-dimensional state-space description in which stability/feedback proximity is governed by the denominator $1 + \Gamma_S \Gamma_L z(\Omega)$, whereas directional channel suppression (CPT) is governed by the numerator constraint on the selected output channel. Importantly, CPT–EP occurs inside a near-unitary window where $|1 + \Gamma_S \Gamma_L z(\Omega_0)|$ stays finite (no resonant pole), in contrast to resonance ($1 + \Gamma_S \Gamma_L z \to 0$) or CPA (singular-value collapse of the full two-port $\mathbf{S}$ matrix).

**S2. Analytical numerator–denominator structure and CPT–EP condition**

Using the ABCD-to-$\mathbf{S}$ expressions above, all $\mathbf{S}_{ij}$ share the same denominator $\text{Den}(\Omega)$, so $b_1(\Omega)$ can be written as a single fraction:

$$b_1(\Omega) = \frac{N(\Omega)}{D(\Omega)}, \tag{S9}$$

where

$$D(\Omega) = \text{Den}(\Omega) = A + B + C + D, \tag{S10}$$
$$N(\Omega) = (A + B - C - D)a_1(\Omega) + 2(AD - BC)a_2(\Omega). \tag{S11}$$

We follow the ABCD cascade:

$$M_{\text{tot}} = M_S\, M_p(\Omega)\, M_I\, M_p(\Omega)\, M_L, \quad M_p(\Omega) = \begin{pmatrix} \cosh K & \sinh K \\ \sinh K & \cosh K \end{pmatrix}. \tag{S12}$$

Each interface matrix $M_j$ ($j \in \{S, I, L\}$) is constructed as a unitary boundary, hence $\det M_j = 1$ and $\det M_p = 1$, implying $AD - BC = 1$, for the total ABCD matrix $M_{\text{tot}} = \begin{pmatrix} A & B \\ C & D \end{pmatrix}$. Using the standard ABCD-to-**S** relations (power-normalized),

$$b_1 = S_{11}a_1 + S_{12}a_2 = \frac{(A + B - C - D)a_1 + 2(AD - BC)a_2}{A + B + C + D}, \tag{S13}$$

we can rewrite the numerator as a quadratic polynomial in $z$,

$$b_1(z) = \frac{N(z)}{D(z)}, \quad N(z) = n_0 + n_1 z + n_2 z^2, \tag{S14}$$

with the explicit coefficients $n_0 = \Gamma_S\, a_1$, $n_1 = -\Gamma_I\,(1 + \Gamma_S\Gamma_L)a_1 - i\,\tau_S\tau_I\tau_L\, a_2$, $n_2 = \Gamma_L\, a_1$. Here $\Gamma_S, \Gamma_I, \Gamma_L$ are the (possibly complex) reflection coefficients of the source/interface/load boundaries used in the code, and $\tau_j = \sqrt{1 - |\Gamma_j|^2}, j \in \{S, I, L\}$, is the corresponding power transmission amplitude. Under the same cascade, the common denominator in Eq. (S14) reduces to

$$D(z) = d_0 + d_1 z + d_2 z^2, \tag{S15}$$

with $d_0 = 1, d_1 = -\Gamma_I(\Gamma_S + \Gamma_L), d_2 = \Gamma_S\Gamma_L$.

Since $A, B, C, D$ are functions of $K(\Omega)$ and the interface parameters ($\Gamma_S, \Gamma_I, \Gamma_L$), and because $z(\Omega) = e^{-2K(\Omega)}$, we may equivalently write $N(\Omega) = N[z(\Omega)]$ and $D(\Omega) =$

$D[z(\Omega)]$. In the present passive cascade, CPT–EP is engineered such that $D(\Omega_0) \neq 0$ (near-unitary window), while $N[z(\Omega)]$ acquires a double zero under a fixed coherent operating input. We fix $\Gamma_S$ (magnitude and phase) and set $\phi_L = \pi$. The three unknowns $(|\Gamma_L|, \Omega_0, |\Gamma_I|)$ are obtained by enforcing two real closure conditions and one critical-coupling condition. First, at $\Omega_0$ we impose the complex closure equation in terms of $z(\Omega)$:

$$z(\Omega_0) = z_{\text{EP}}(\Gamma_S, \Gamma_L). \tag{S16}$$

The target $z_{\text{EP}}$ is given in closed form as

$$z_{\text{EP}} = -\sqrt{\frac{\Gamma_S}{\Gamma_L}}, \tag{S17}$$

so we enforce $\text{Re}[z(\Omega_0) - z_{\text{EP}}] = 0$ and $\text{Im}[z(\Omega_0) - z_{\text{EP}}] = 0$.

Second, the internal interface magnitude is set to the EP-critical value

$$|\Gamma_I|_{\text{EP}} = \frac{2\sqrt{|\Gamma_S \Gamma_L|}}{1 + |\Gamma_S \Gamma_L|}. \tag{S18}$$

To ensure the CPT–EP is not a resonance-induced pole, we require the common denominator to remain bounded at $\Omega_0$, i.e., $D[z(\Omega_0)] \neq 0$, equivalently $|1 + \Gamma_S \Gamma_L z(\Omega_0)| \geq \epsilon_D$, where $\epsilon_D$ is a small numerical threshold (we use $\epsilon_D = 10^{-3}$ unless stated otherwise). Operationally, the near-unitary window is verified by $\sigma_{\max}(\Omega) \approx 1$ and $\sigma_{\min}(\Omega) \approx 1$ within a prescribed tolerance, while the CPT channel is suppressed by $|b_1(\Omega)| \ll 1$.

**S3. Coherent-input protocols, quartic leakage derivation**

Two protocols are employed: (i) per-frequency probe $\mathbf{a}_{\text{pf}}(\Omega)$ that nulls $b_1(\Omega)$ identically:

$$a_{1,\text{pf}}(\Omega) = -S_{12}(\Omega), a_{2,\text{pf}}(\Omega) = S_{11}(\Omega), \tag{S19}$$

followed by normalization $\|\mathbf{a}_{\text{pf}}\| = 1$, giving

$$b_{1,\text{pf}}(\Omega) = S_{11}a_{1,\text{pf}} + S_{12}a_{2,\text{pf}} \equiv 0. \tag{S20}$$

(ii) fixed coherent input defined at the EP frequency and held constant:

$$\mathbf{a}_{\text{fixed}} = \frac{\mathbf{a}_{\text{pf}}(\Omega_0)}{\|\mathbf{a}_{\text{pf}}(\Omega_0)\|}. \tag{S21}$$

The per-frequency probe transfers sensitivity into the required input ratio and phase $a_2/a_1 = -S_{11}/S_{12}$, whereas the fixed protocol reveals the intrinsic output scaling. By construction, $b_{1,\text{fixed}}(\Omega_0) = 0$ because $\mathbf{a}_{\text{fixed}}$ is chosen from the nulling vector at $\Omega_0$. For $\Omega$ near $\Omega_0$, write $b_{1,\text{fixed}}(\Omega) = N(\Omega)/D(\Omega)$ with $N, D$ defined in (S10)–(S11). In the near-unitary window $D(\Omega_0) \neq 0$, so the leading behavior is governed by the numerator expansion. Using $z(\Omega) = e^{-2K(\Omega)}$ and the EP tuning (S17)–(S18), the numerator acquires a double root at $\Omega_0$:

$$N[z(\Omega_0)] = 0, \frac{dN}{dz}\bigg|_{z(\Omega_0)} = 0. \tag{S22}$$

Therefore, the first non-vanishing term in the complex amplitude is quadratic:

$$b_{1,\text{fixed}}(\Omega) = c(\Omega - \Omega_0)^2 + \mathcal{O}[(\Omega - \Omega_0)^3], \tag{S23}$$

where the quadratic coefficient is

$$c = \frac{1}{2}\frac{d^2 b_{1,\text{fixed}}}{d\Omega^2}\bigg|_{\Omega=\Omega_0} = \frac{1}{2}\frac{N''(z_0)}{D(z_0)}\left(\frac{dz}{d\Omega}\bigg|_{\Omega_0}\right)^2, \tag{S24}$$

with $z_0 = z(\Omega_0)$ and $N''(z_0) = \frac{d^2 N}{dz^2}\big|_{z_0}$. The leaked power obeys the quartic law

$$P_{1,\text{fixed}}(\Omega) = |b_{1,\text{fixed}}(\Omega)|^2 \propto |\Omega - \Omega_0|^4. \tag{S25}$$

In contrast, $b_{1,\text{pf}}(\Omega) = 0$ by construction, so no intrinsic output scaling can be extracted from $P_{1,\text{pf}}$; EP sensitivity is instead reflected in the frequency dependence of the required coherent input.